      [1999/12/01 v1.4c Il Nuovo Cimento]
\documentclass[preprint]{cimento}

\usepackage{latexsym}
\usepackage{graphics}
\usepackage{graphicx}
\usepackage{bm}
\usepackage{amsmath}
\usepackage{amsfonts}
\usepackage{amssymb}
\usepackage{epsfig,array}

\def\beqa{\begin{equation}}
\def\eeqa{\end{equation}}
\def\bea{\begin{array}}
\def\eea{\end{array}}
\def\beq{\begin{equation}}
\def\eeq{\end{equation}}
\def\ba{\begin{eqnarray}}
\def\ea{\end{eqnarray}}
\def\gaga{\gamma\gamma}

\title{{$\gaga \to \mu \tau b \bar{b}$} in
Susy Higgs mediated lepton flavor violation
\thanks{Prepared for \emph{Incontri di Fisica delle Alte Energie}, IFAE
2009, Bari, Italy, 15-17 Aprile 2009. To appear in the proceedings to be published by the Societ\`{a} Italiana di Fisica  in the  Nuovo Cimento C - Colloquia on Physics }
}

\author{
Mirco Cannoni \from{ins:x}
\atque
Orlando~Panella \from{ins:evil}}
\instlist{\inst{ins:x} Universit\`a di Perugia, Dipartimento di Fisica,
Via A.~Pascoli, 06123, Perugia, Italy.
\inst{ins:evil} INFN, Sezione di Perugia, Via A.~Pascoli, 06123 Perugia, Italy}

\begin{document}

\maketitle

\begin{abstract}
The process $\gaga\to\mu\tau b\bar{b}$ is studied in the minimal supersymmetric standard model within a large $\tan\beta$ scenario
imposing on the parameter space
present direct and indirect constraints from $B$ physics and rare LFV $\tau$-decays.
At a photon collider based on an $e^+e^-$ linear collider with $\sqrt{s}=800$ GeV with the parameters of the TESLA proposal (expected integrated $\gamma\gamma$-luminosity $L_{\gamma\gamma}=200\div 500$ fb$^{-1}$)
the LFV signal can be probed for masses of the heavy neutral Higgs bosons
$A,H$ from $300$ GeV up to the kinematical limit $\simeq 600$ GeV for 30$\leq\tan\beta\leq$60.
\end{abstract}

The minimal super symmetric standard model (MSSM) (like the standard model)
does not provide any explanation for the neutrino
masses and mixing. In order to accomplish this task, the seesaw
mechanism is usually implemented in the MSSM adding
right handed neutrinos ($\nu$-MSSM).
Compared to the
MSSM, the main novelty in the $\nu$-MSSM is the presence of lepton
flavor violation (LFV). LFV effects arise both in the
gauge interactions~\cite{borzumati} (through lepton-slepton-gaugino couplings)
and in the Yukawa interactions~\cite{bkl}.
In particular, LFV Yukawa interactions are greatly enhanced at large $\tan\beta$, and give
the possibility of detecting LFV decays of the Higgs bosons at LHC~\cite{brignole1,moretti} and ILC in the $e^+e^-$ mode~\cite{kanemura1}.
In Refs. \cite{Cannoni} loop level lepton flavor violating processes such as $e^+e^- \to e^+\ell^-$ ($\ell=\mu\tau$),
and $\gamma\gamma\to \ell_i\ell_j$ ($\ell_i \neq \ell_j$),
which are potentially striking signatures of LFV, were studied in detail.
Here we discuss a new mechanism of lepton flavor violation at the photon collider
\cite{ILC,ginzburg} via the Higgs mediated ($H,A$) process:
$\gamma \gamma \to \mu \tau b \bar{b}$
in a  scenario of large $\tan\beta$ where all the super-partner masses are $\cal{O}$(TeV) and
the heavy Higgs bosons ($A,H$) have instead masses below the TeV and develop sizable
loop induced LFV couplings to the SM leptons.

\begin{figure}[t!]
\begin{center}\hspace{1.0cm}
\includegraphics[scale=0.37]{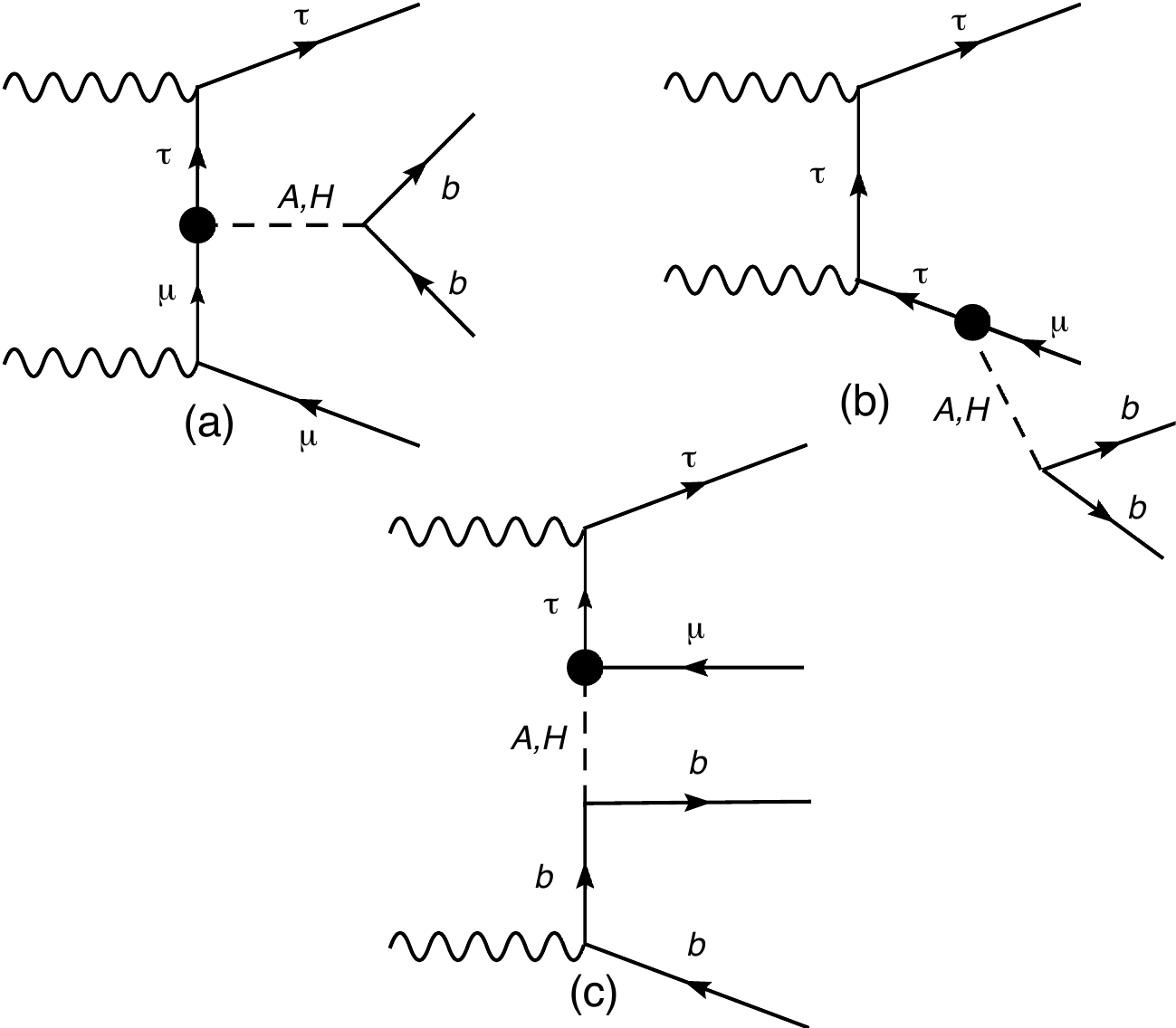}\hspace{0.1cm}
\includegraphics[scale=0.7]{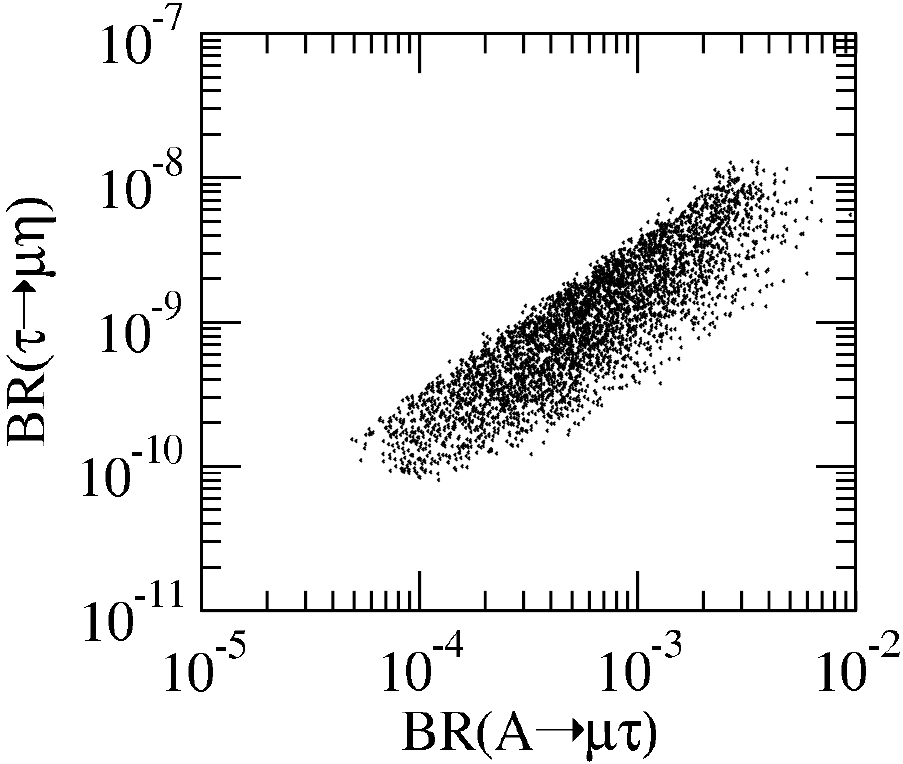}
\includegraphics[scale=0.7]{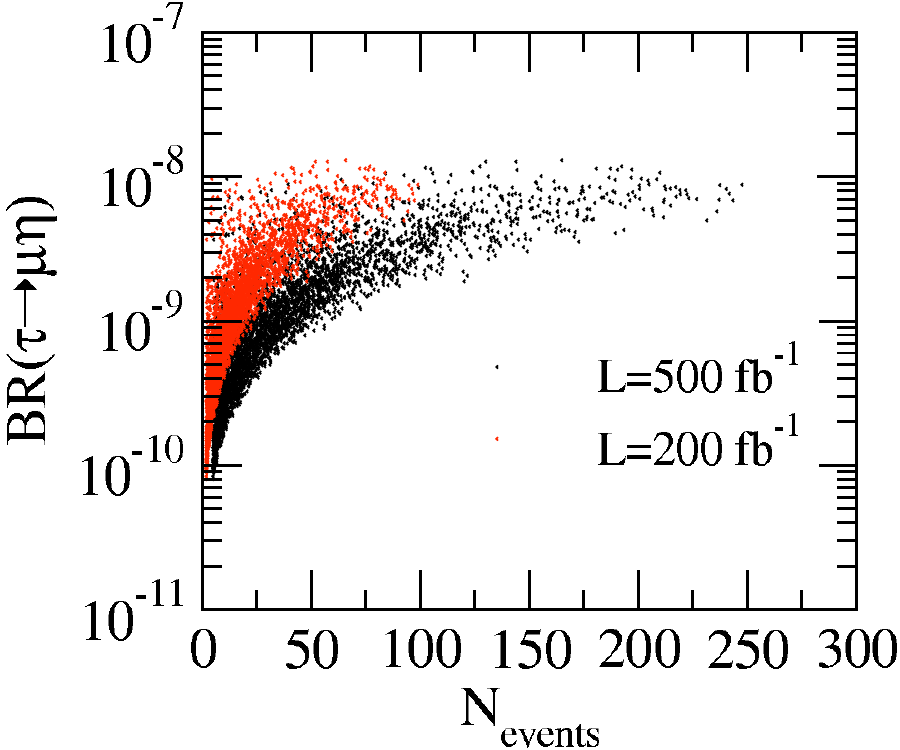}
\includegraphics[scale=0.7]{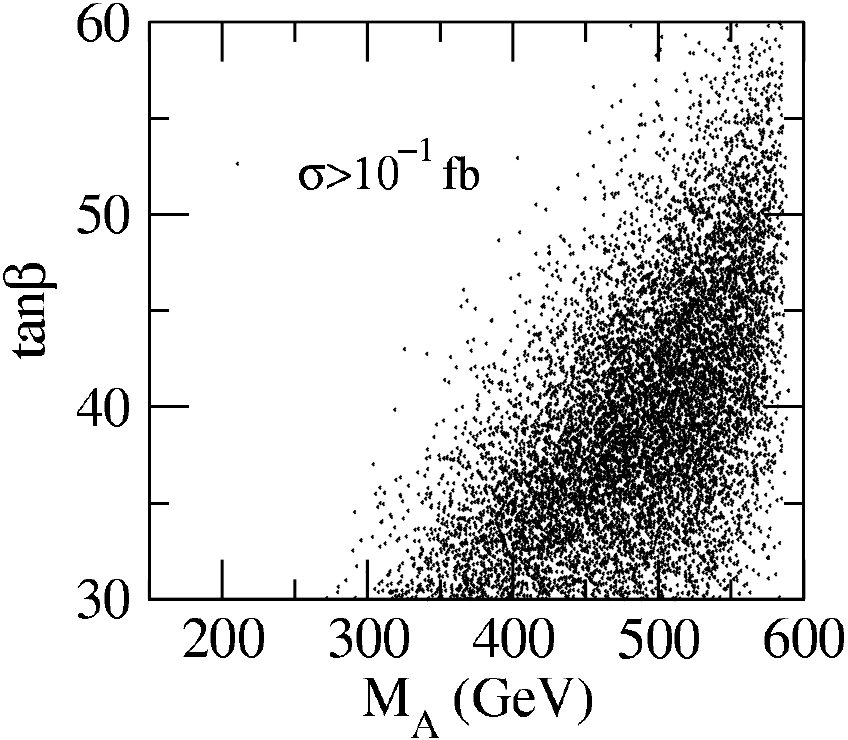}
\caption{(Top left panel) Diagrams for the process $\gamma\gamma \to \mu\tau
b\bar{b}$: the topology (a) is the one we call $\mu\tau$ fusion. The
black blob represents the loop induced LFV coupling treated as an
effective vertex.
(Top right  panel) Correlation between  ${\cal
B}(\tau\to\mu\eta)$ and $A \to \mu \tau$.
(Bottom left panel) Correlation between ${\cal B}(\tau\to\mu\eta)$ and
the number of expected events for two values of the
integrated luminosity.
(Bottom right panel)
Distribution of
the signal cross section in the ($M_A , \tan\beta$) plane.
The parameter space and the imposed constraints are discussed in the text.
}
\label{fig}
\end{center}
\end{figure}

In photon-photon collisions the main production mechanisms
for the Higgs bosons are $\gamma\gamma$ fusion and
$\tau\tau$ fusion~\cite{choi}. In the first case, the Higgs
is produced as an s-channel resonance through a loop
involving the exchange of massive charged particles. In the
$\tau\tau$ fusion process $\gamma\gamma \to \tau\tau
b\bar{b}$, the Higgs is produced in the s-channel with a
$\tau\tau$ pair and can be detected with its decay mode
$b\bar{b}$. We have shown in Ref.~\cite{cannoni} that the
main LFV process is the $\mu\tau$ fusion to the Higgs, see
diagram (a) in Fig.~\ref{fig}, top-left panel, which
dominate the $\gaga$ fusion, with large cross section over
large portion of the parameter space. The signal from the
$\mu\tau$ fusion $\gaga \to \mu\tau b\bar{b}$ consists of a
$\mu\tau$ pair plus $b\bar{b}$ jets from the Higgs decay,
allowing the possibility to detect and reconstruct the
Higgs through its main decay channel and to measure, at the
same time, the size of LFV couplings. In the following we
report the main results while detailed analysis of the
signal and background can be found in Ref.~\cite{cannoni}.

In the mass-eigenstates basis for both leptons and Higgs bosons,
the effective flavor-violating Yukawa interactions are described by the
lagrangian:
\beqa
\mathcal{-L}\simeq(2G_{F}^2)^{\frac{1}{4}}\frac{m_{l_i}} {c^2_{\beta}}
\left(\Delta^{ij}_{L}\overline{l}^i_R l^j_L+\Delta^{ij}_{R}
\overline{l}^i_L l^j_R \right)
\times \left(c_{\beta-\alpha}h-s_{\beta-\alpha}H-iA \right) +h.c.
\label{lagrangian}
\eeqa
where $\alpha$ is the mixing angle between the CP-even Higgs bosons $h$ and
$H$, $A$ is the physical CP-odd boson,
and  we adopt the notation
$(c_{\theta},s_{\theta},t_{\theta})\! =\!(\cos \theta,\sin \theta,\tan \theta)$. In
Eq.~(\ref{lagrangian}) $i,j$ are flavor indices that in the following are understood to be
different ($i\neq j$).

The couplings $\Delta^{ij}$ in Eq.~(\ref{lagrangian}) are
induced at one loop level by the exchange of gauginos and
sleptons, provided a source of slepton mixing is present.
In this work the analysis at Higgs LFV effects will be
model independent and we use the expressions of
$\Delta^{ij}_{L,R}$ obtained in the mass insertion
approximation. The LFV mass insertions $\delta^{ij}_{LL}$
and $\delta^{ij}_{RR}$ are defined as:
$\delta^{ij}_{LL}\!=\!{({m}^2_{L})^{ij}}/{m^{2}_{L}}$,
$\delta^{ij}_{RR}\!=\!{({m}^2_{R})^{ij}}/{m^{2}_{R}}$,
where $({m}^2_{L})^{ij}$ are the off-diagonal flavor
changing entries of the slepton mass matrix.

The Higgs boson decay widths
relevant for the following analysis at a photon collider are
obtained by means of the lagrangian of Eq.~(\ref{lagrangian}) using the approximation $1/{c}^2_\beta
\simeq \tan^2\beta$ (only valid in the large $\tan\beta$ regime) and in the limit of  massless fermions.
Introducing  $\Delta^2 =|\Delta_L^{32}|^2 + |\Delta_R^{32}|^2$ we find
$\Gamma({A\to \tau^+\mu^-})+\Gamma({A\to\tau^-\mu^+})
=t^{2}_{\beta} \Delta^2 \Gamma({A\to \tau^+\tau^-})$.
For the heavy Higgs boson $H$, the right hand side of the previous equation
should be multiplied by a factor
$({s}_{\beta-\alpha}/{c}_\alpha)^2$.
The light Higgs field $h$ has negligible lepton flavor violating decays since its coupling
$\cos(\beta-\alpha)\to 0$ in the decoupling regime.

In Ref.~\cite{cannoni} we have shown that a relieable
estimate of the cross section, including cuts for
background suppression, is given by considering only the
diagram (a) of Fig.~\ref{fig} (top-left panel) with the
method of the equivalent particle approximation (EPA). The
cross section for monochromatic photons is given by the
convolution of the photon splitting functions to a pair of
leptons~\cite{choi}, with the cross section in the center
of mass frame of the sub-process $\mu\tau \to b\bar{b}$ in
the small width approximation (SWA): \beqa \sigma(\gaga \to
\mu\tau b\bar{b})= \frac{4 \pi^2}{s_{\gaga}}
\frac{\Gamma(A\to \tau\mu) {\cal B}(A\to b\bar{b})}{M_A}
2\int_{-\ln 1/t}^{+\ln 1/t} {d\eta}\; P_{\gamma/ \mu}\left(t e^\eta \right)
P_{\gamma /\tau}\left(t e^{-\eta}\right)
 \label{fusionfinal}
 \eeqa
where $s_{\gaga}$ is the photons center of mass energy
squared, the factor two is the multiplicity factor which
accounts for the exchange of the initial photons, $\eta =
\ln\sqrt{{x_\mu}/{x_\tau}}$, $t={M_A}/{2E_\gamma} $,
$x_{\mu}$, $x_{\tau}$ being the colliding photon's energy
fraction carried by the $\mu$ and $\tau$ .

In order to provide a detailed study of the possibilities
of a photon collider with respect to the LFV violating
signal $\gamma \gamma \to \mu\tau b \bar{b}$ we performed a
scan over the following parameter space: $1\, \hbox{TeV}
\leq (\mu, m_{\tilde{q}}, A_u, A_d, m_{{L}},m_{{R}})\leq5\,
\hbox{TeV}, 500\,  \hbox{GeV} \leq(M_1 , M_2, M_3)\leq 5\,
\hbox{TeV}, 150\, \hbox{GeV} \leq M_A \leq1\, \hbox{TeV},
30\, \leq \tan\beta \leq\, 60, 10^{-3} \leq \,
(\delta_{LL}^{32},\delta_{RR}^{32}) \leq0.5$.

We impose the following constraints on the parameter space:
lower bound on the light Higgs mass $m_h >114.4$ GeV; upper
bound on the anomaly of the muon magnetic moment
$(g-2)_{\mu}< 5\times 10^{-9}~$; bounds on electro-weak
precision observables such as $\Delta\rho<1.5\times
10^{-3}$; direct search constraints on the lightest
chargino and sfermion masses and constrains on squarks and
gluino masses from LEP and Tevatron are automatically
satisfied as they lie in the TeV range in our scenario.
Some $B$-physics processes, namely $B_s \to \mu^+ \mu^-$,
$B \to X_s \gamma$ and $B_u \to \tau \nu$, are particularly
sensitive to $\tan\beta$ \cite{bkq}. We require that the
parameter space satisfies ${\cal B}(B_s \to \mu^+
\mu^-)<6.5\times 10^{-8}$ \cite{pdg}; $R_{\tau\nu}$, the
ratio between the SUSY and SM branching ratios for $B_u \to
\tau \nu$, is required in the range
$0.70<R_{\tau\nu}<1.44$; $R_{X_s\gamma}$, the ratio between
the SUSY and SM branching ratios for $B \to X_s \gamma$, is
required to lie in the range $1.01<R_{X_s\gamma}<1.25$
\cite{masiero3}. We impose the current  upper bounds on LFV
$\tau$ decays to be respected: ${\cal B}(\tau^- \to \mu^-
\eta)<6.8\times 10^{-8}$ and ${\cal B}(\tau^- \to \mu^-
\gamma)<5.6\times 10^{-8}$~\cite{pdg}. In the case where
Higgs-mediated LFV effects are important, $\tau\to\mu\eta$
is generally the dominant process~\cite{bkq}.

In the top-right panel we show the correlation
between ${\cal B}(\tau\to\mu\eta)$ and ${\cal
B}(A\to\mu\tau)$. The latter gets values in the interval $(5\times10^{-4})\lesssim {\cal
B}(A\to\mu\tau) \lesssim (8\times 10^{-3})$. Even if
the upper limit on  ${\cal B}(\tau\to\mu\eta)$ is lowered by an order of
magnitude from its actual value ($\approx 10^{-8}$) we see that ${\cal
B}(A\to\mu\tau)$  can still reach values up to ${\cal O}(10^{-3})$ which are
particularly interesting for the  LHC, where
the cross section for heavy neutral gauge bosons production
in $b\bar{b}$ fusion is sizable~\cite{moretti}.

In the bottom-left panel of Fig.~\ref{fig} we show the
correlation between ${\cal B}(\tau\to\mu\eta)$ and the
number of $\mu\tau b\bar{b}$ events corresponding to the
cross sections for monochromatic photon collisions at
$\sqrt{s_{\gaga}}=600$ GeV for two values of the integrated
luminosity, ${\cal L}=200-500$ fb$^{-1}$/yr. It can be seen
that for the high luminosity option we can expect up to 250
events per year, and up to 100 events per year for the low
luminosity option. The above conclusions are valid for the
present upper limits on the branching ratios.

In the bottom-right panel we show the region of the
parameter space in the $(M_A ,\tan\beta)$ plane which is
characterized by a signal cross-section $\sigma \geq
10^{-1}$ fb. The signal cross-section becomes larger at low
$M_A$ masses, but in the considered region of large
$\tan\beta$ values such low masses are excluded by the
imposed constraints. In particular, the LFV signal for
$M_A$ masses below $300$ GeV are excluded for all values of
$\tan\beta$ in the interval, $30 <\tan\beta<60$. We have
checked that requiring a signal cross section $10^{-2}
\text{fb} \leq \sigma \leq 10^{-1}$ fb the same region in
the $(M_A ,\tan\beta)$ plane is covered.

\end{document}